\documentclass[12pt]{article}
\usepackage{epsfig}

\def\a{\kern+.6ex\lower.42ex\hbox{$\scriptstyle \iota$}\kern-1.20ex a}
\def\e{\kern+.5ex\lower.42ex\hbox{$\scriptstyle \iota$}\kern-1.10ex e}

\def\A{\kern+.6ex\lower.42ex\hbox{$\scriptstyle \iota$}\kern-1.20ex A}
\def\E{\kern+.5ex\lower.42ex\hbox{$\scriptstyle \iota$}\kern-1.10ex E}
\newcommand{\nc}{{\cal N}}
\begin{document}
\title{Studying thermodynamics
 in heavy ion collisions}
\author{ A.Bialas, W.Czyz  and J.Wosiek\\ M.Smoluchowski Institute of
Physics\\Jagellonian University, Cracow\thanks{address: Reymonta 4, 30-059
Krakow, Poland; e-mail:wosiek@thp4.if.uj.edu.pl}}
\maketitle
\begin{abstract}
We discuss  the possibility to measure entropy of the system created in heavy
ion collisions using the  Ma coincidence method.
\end{abstract}

\vspace{0.3cm}
\section{Introduction}
\vspace{0.3cm}

The assumption of thermodynamic equilibrium is one of the most commonly used
when discussing the system created in central collisions of two relativistic
nuclei. It is by no means obvious, however, that the
 equilibration actually can be achieved, since it is recognized as a process
which
may take longer time than the life time of the system in question. Be
it or not, it is certainly important to verify if the created system is
indeed in thermal equilibrium. To test this, one needs to check if the
various measured quantities do satisfy the relations
following from thermodynamics. In the present note we discuss the possibility
of testing the relations \cite{la1}
\begin{equation}
\left. \frac{\partial S(E,n)}{\partial E}\right|_{n} = \frac1{T},   \label{1.1}
\end{equation}
and, with $\mu$ denoting the chemical potential,
\begin{equation}
\left. \frac{\partial S(E,n)}{\partial n}\right|_{E}=-\frac{\mu}{T}, \label{1.2}
\end{equation} 
which should be valid in any system at thermal equilibrium.

Testing (\ref{1.1}) requires measurement of the temperature $T$, the energy
$E$,
the number of particles $n$ 
and the entropy $S$ of the system in question. It is clear that by measuring
the
energies of the particles created in the collision we can measure the energy of
the system. It is also generally accepted that by measuring the slope in
the transverse momentum distribution
we can measure the temperature\footnote{This requires correction to the
effects of the hydrodynamic flow which seem to be under a reasonable control.}.
The  real difficulty is the measurement of entropy. In the present note we
propose to adapt to this end the concidence method proposed some time ago by
Ma \cite{ma1}. We also present Monte Carlo estimates of the feasibility of the
method, based on a simple model. 
We conclude that the method has a large
potential, as it requires much smaller number of events 
($\sim \sqrt{no.\; of\;
states}$), then the conventional approach. As a consequence its errors are
significantly reduced compared to the simple-minded estimates. It is
certainly worth to employ it in the present and future high energy
experiments. 

\section{Measurement of entropy by the coincidence method}
Below we give a summary of the
idea presented in \cite{ma1}.

Ma proposes to count the pairs of configurations of the investigated system.
Call $N_c$ the number of pairs of  "identical" configurations. Call $N_t$ the
total
number of pairs of configurations. If all configurations considered are
"equivalent"
(i.e. if they correspond to the same conditions), then entropy is given by
the formula
\begin{equation}
S=\log\left(\frac{N_t}{N_c}\right) .    \label{2.1}
\end{equation}
The reason is that $\frac{N_t}{N_c}$ is the volume in the phase-space occupied
by the system. This can be seen as follows.

Suppose that the phase-space is divided in cells. Suppose furthermore that our
system occupies $\Gamma$ cells (with uniform probability). Each cell represents a
different state of the system (each cell has as many dimensions as is the
number of variables describing the system). Our problem is to calculate
$\Gamma$: $S=\log(\Gamma)$. Let us select randomly
$\nc$ configurations of the system (in general $\nc \ll \Gamma$, that's the main
point). These configurations occupy some cells. 
The average occupation number of a cell is $\nc/\Gamma\ll 1$. 
Under this condition, the avergae number of pairs in 
the same cell 
is
\begin{equation}
\left(\frac{\nc}{\Gamma}\right)^2\approx \frac{N_t}{\Gamma^2},   \label{2.1a}
\end{equation}
where $N_t\approx \nc^2$ is the total number of pairs selected. The total
number of coincidences is the sum of (\ref{2.1a}) over all cells
\begin{equation}
 N_c=\Gamma \left(\frac{\nc}{\Gamma}\right)^2,          \label{2.1b}
\end{equation}
hence
\begin{equation}
 \Gamma =\frac{N_t}{N_c},            \label{2.1c}
\end{equation}
and thus (\ref{2.1}).\footnote{The formula (\ref{2.1a}) is only approximate.
The exact formula is
$ \frac{\nc}{\Gamma}\frac{\nc-1}{\Gamma}$ which leads again to (\ref{2.1c}).}

If the configurations are not equivalent, one has to divide them into classes:
within each class they are now equivalent. If the probability distribution of
classes is $P(\lambda)$, then
\begin{equation}
S=\sum_{\lambda} P(\lambda)
\log\left(\frac{N_t(\lambda)}{N_c(\lambda)P(\lambda)}\right).   \label{2.2}
\end{equation}
The derivation is given in \cite{ma1} but it can be easily understood as a sum
of the "average over classes"
$=\sum_{\lambda} P(\lambda)
\log\left(\frac{N_t(\lambda)}{N_c(\lambda)}\right)$ and of the "entropy of the
distribution of classes"
$=-\sum_{\lambda} P(\lambda)\log[P(\lambda)]$.

The classes cannot be too small, so that number of configurations in each class
is sufficient to make a reasonable statistics.

This is what we retain from Ma.  In the next
section we present a  suggestion how to apply this method to measure the
entropy of a system of particles produced in high energy interactions.

\section{Application of coincidence method to multiparticle production}

A natural possibility to apply the coincidence method
to multiparticle production is to identify the configurations of the
statistical system with the events observed in experiment.
Once this is accepted, one can proceed as follows.

(a) Select $\nc$ events and split them into classes according to the total
transverse energy $E$ and multiplicity $n$ recorded. The number of events in each
class is denoted by $\nc(E,n)$. 

(b) Define a "lattice" in momentum space (e.g. rapidity, azimuth  and $\log
p_t, or \log E_t$) of individual particles \footnote{Probably the better
method is to transform the momenta into variables which give uniform
distributions (see e.g. \cite{bg}).}.

 Within each class:

(i) Call the two configurations "identical" if they have the same occupation
numbers within the accuracy of the grid. The number of such pairs is denoted by
$N_c(E,n)$.

(ii) Calculate the ratio (\ref{2.2}), i.e.,
\begin{equation}
S(E,n)=
\log\left(\frac{N_t(E,n )}{N_c(E,n)}\right)=
\log\left(\frac{\nc(E,n)(\nc(E,n)-1)}{N_c(E,n)}\right)    \label{2.3}
\end{equation}
where $\nc(E,n)$ is the number of events in a given class.

Actually,  the condition that
 the events in one
"equivalence class" must have strictly the same multiplicity (otherwise they
could never be really identical)  could be relaxed, e.g. by accepting that in
the definition of the
"identity" of the two configurations, the occupation numbers may differ by a
small amount.

If this procedure is going to have any sense, the results should depend
on
the lattice spacing $a$ in a trivial way: when $a\rightarrow a'$,
$S\rightarrow S'= S+\log(a/a')$ (once the spacing is small enough). This must
be checked, of course.

 Also question of size of the energy bins
must be analysed. Again, the result should  depend on the chosen size of
the energy bins in a trivial way: if the size of the energy bin $\Delta$
changes into $\Delta'$ the entropy $S$ changes into
$S'=S+\log(\Delta/\Delta')$.
 But this is delicate: the bins should be small enough so that
within a bin the energy may be considered constant but large enough so that
there is a reasonable statistics within each bin. Same applies to the multiplicity.

One sees  from these arguments that in this way one can measure the entropy
only up to an additive constant. Therefore the interesting thing is not to
measure the absolute value of entropy but rather its dependence
on energy or multiplicity.

As we already noted in the Introduction,
the measurement described by (\ref{2.3}) 
allows one to perform a simple test
of thermalization. When thermodynamics is valid, then the
Eqs.(\ref{1.1},\ref{1.2}) should
be satisfied. Clearly the additive constant is irrelevant. One needs, however,
a rather precise measurements because otherwise the numerical estimates of the
derivatives $ \frac{\partial S(E,n)}{\partial E}$ and 
 $ \frac{\partial S(E,n)}{\partial n}$ are not reliable. In the next section we show
the results of a simple Monte Carlo estimate of the accuracies one can achieve.

\section{The classical gas of identical particles}

Of course the main problem is for how big systems the coincidence method 
works in practice. The number of states grows exponentially with the 
number of particles and the number of subdivisions. Therefore obviously
there is a limit to what one can achieve with the finite computer. We will
show here however that, the onset of the thermodynamic behaviour occurs  
still for the sizes where the coincidence method is feasible
practically. 

Second, we will also show what must be the {\em minimal} size of the system
for the continuum behaviour to set in. For very small number of subdivisions
the coincidence method and our Monte Carlo are quickly convergent, however
they converge to the correct description of the discrete problem, which is
far from the continuum, hence not interesting.

We consider the classical gas of noninteracing, nonrelativistic particles in
$d$ dimensions. Since, as mentioned earlier, the method may be applied to
the transverse degrees of freedom only we prefer to retain the discussion
in arbitrary dimensions. For the same reason we will use the number of
degrees of freedom $N$ to characterise the size of a system. Of course
$N=d\; n$ in $d$ dimensions \footnote{Provided only the momentum
degrees of freedom are considered.}.

Since for noninteracting particles momentum and space degrees of freedom factorize,
we consider for simplicity only the momentum states. The discretized expression for 
the number of states of $N$ degrees of freedom with the total energy E reads
\begin{equation}
\Gamma(M,N)=\sum_{n_1,\dots n_N, n_1^2+\dots n_N^2=M} 1,
\end{equation}
where the momentum $p_i=a n_i$ with some discretization scale $a$.
Accordingly $2 m E = a^2 M $ , where the integer $M$ labels the energy of the system
and $m$ denotes the mass of a particle. 

The generating function
\begin{equation}
Z_N(\tau)=\sum_{M=0}^{\infty} \tau^M \Gamma(M,N)=\left(\sum_{n=-\infty}^{\infty} \tau^{n^2}
\right)^N = \exp{(N \log {c(\tau)})}, \label{zet} 
\end{equation}
factorizes and is expressed by a single sum $c(\tau)$. The coefficients $\Gamma$ can now be 
simply obtained by calculating recursively expansion of the $f(\tau)=\log{c(\tau)}$ from
that of $c(\tau)$, and subsequently expansion of $Z_N(\tau)$ from that of $f(\tau)$. 
This procedure provided us with the exact numbers for the density of states $\Gamma(M,N)$,
which were used to benchmark the performance of our Monte Carlo.

For large $M$ and $N$ the density of states reaches its continuum limit
\begin{equation}
\Gamma(M,N) \cong  \frac{\pi^{N/2} M^{(N/2-1)}}{(N/2-1)!},\;\;M,N\;large.  \label{cont}
\end{equation} 
We will see later that this relation is rather well satisfied even for moderate
values of $M$ and $N$.

On the other hand, {\em the thermodynamic} limit, $M,N \rightarrow \infty, M/N$-fixed, 
is reached rather slowly. In this limit the entropy density \footnote{Here and in the 
following, we will refer to the entropy per one degree of freedom as the entropy density.}
scales depending only on the energy density $\epsilon=M/N$.
\begin{equation}
 \frac{1}{N}\log{\Gamma(M,N)}\cong \frac{1}{2}(\log{(\epsilon)}+\log{(2\pi)}+1),\;\;
M, N \rightarrow \infty, \epsilon=M/N - const. \label{thermo}
\end{equation}
 The purpose of this exercise is to see if the
 coincidence method can detect this behaviour.
\subsection{Monte Carlo simulations}
  In principle one should generate a sample of $\nc$
 {\em configurations} $\{n_1,n_2,\dots,n_N\}_k$, $k=1,...,\nc$ of, integer-valued,
 one-dimensional momenta $n_1,\dots,n_N$, which satisfy the energy conservation.
 For our purpose, however, the details of particle kinematics, 
although practically
cumbersome, are not relevant. In order to measure the
 coincidences, it is enough to
 label  uniquely all multiparticle states and compare the labels. In this way the problem
simplifies considerably, yet the essential question of the onset of the thermodynamic 
behaviour can be addressed.

    Consequently each Monte Carlo run consisted of a generation of a sample 
of $\nc$ configurations, represented by integer indices, $(I_1,I_2,{\dots},I_{\nc}),
  1\leq I_k \leq \Gamma(M,N), k=1,\dots ,\nc$, uniformly distributed in the whole
space of available states. Then we counted the number of coincidences 
$\hat{N}_c$, i.e., the number of pairs $(I_j,I_k)$ such that $I_j=I_k$. The estimate for the number
of all states is then
\begin{equation}
\hat{\Gamma}=\nc(\nc -1)/\hat{N}_c.  \label{esti}
\end{equation}
Moreover assuming the multinomial distribution of $\nc$ integers among $\Gamma$ bins
we have calculated also the higher moments of the distribution of the number
 of coincidencies $N_c$.
In particular, the dispersion of $N_c$ reads \footnote{After some approximations valid 
for $1 \ll \nc  \ll \Gamma$ }
\begin{equation}
\sigma^2[N_c]=2<N_c>=\frac{2\nc^2}{\Gamma},  \label{sigma}
\end{equation}
which gives for the relative error of the determination of $\Gamma$ after $\nc$ trials
\begin{equation}
\frac{\sqrt{\sigma^2[\Gamma]}}{\Gamma}=\frac{\sqrt{2\Gamma}}{\nc}. \label{erth}
\end{equation}
Therefore the estimate of the error, based on the MC data only, is
\begin{equation}
\hat{\sigma}[\Gamma]/\hat{\Gamma}=\sqrt{2/\hat{N}_c}. \label{err}
\end{equation}
Eqs.(\ref{esti},\ref{erth}) show directly another advantage of the coincidence method.
 Namely, it works
for much smaller number of trials $(\sim \sqrt{\Gamma})$ than the standard approach
 which measures average occupation of a single state.

A sample of runs is summarized in Table 1.  Exact results for $\Gamma(M,N)$ are also quoted.
The last column gives the relative deviation of the current estimate (col. 5) from the 
exact value. It should be compared with the estimate of the error based only on the 
Monte Carlo data, Eq.(\ref{err}), given in column 6. The estimated error is steadily
deacresing like $1/\nc$ and actual deviation follows the suit albeit with some fluctuations.
In all runs we have made (about 20 times more than shown in the Table) approximalety 30\% 
of actual deviations were bigger that the MC estimate, as they should.
Of course the formula (\ref{erth}) is essential for planning future Monte Carlo simulations. 
 It is interesting to note that the errors decrease as a 
number
of trials and not as $1/\sqrt{\nc}$. This is because
 the true random variable in this problem is the number of pairs, i.e. ${\nc}^2$. In 
particular the computing effort (counting pairs) grows like ${\nc}^2$, and consequently the
square root of the computational effort determines decrease of errors as it should. 
Altogether the Monte Carlo results are well under contol
and show that the method is quite reliable. It is however practical only if the total number
of states is less than several hundred milions. The last run shown in Table 1 lasted
few hours on a 200 MHz PC. This translates into $N, M \leq\sim 25 $. We will discuss now if 
this is sufficient to see the onset of themodynamic properties.
\subsection{Results}
     Figure 1 shows the entropy density as a function of a scaling variable $\epsilon=M/N$.
Statistical errors of MC results (and the deviation from the exact discrete values given by 
$\Gamma(M,N)$) are much smaller that the size of symbols. The data follow nicely the curves
obtained from the classical formula in the continuum, Eq.(\ref{cont}). Considered as a 
function of $\epsilon$ and $N$ they obviously show a substantial $N$-dependence. 
The $N$ varies from 8 (lowest curve) to 24 in
 this plot. On the other hand, the deviation from the ultimate scaling limit, 
(Eq.(\ref{thermo}), the uppermost curve), is around 30\% in the worst case (N=8,M=30).
With $N$ starting from 12, deviations from the infinite system are smaller than 20\%.
Note that $N$ denotes the nubmer of degrees of freedom, which in $d$ space dimensions 
 corresponds to $N/d$ particles.  

As a second test we have checked a differential form of Eq.(\ref{thermo})
\begin{equation}
\frac{\partial \log{\Gamma}}{\partial E}= \frac{N}{2E},
\end{equation}
which, together with the equipartition of enegry, is the basis of the equilibrium
thermodynamics.
 Changing the variable $2 m E = a^2 M$ gives
\begin{equation}
\frac{\partial \log{\Gamma}}{\partial E}=\frac{\partial\log{\Gamma}}{\partial M}
\frac{d M}{d E}=
\frac{\partial\log{\Gamma}}{\partial M}\frac{M}{E}=\frac{N}{2E},
\end{equation}
or 
\begin{equation}
\frac{\partial\log{\Gamma}}{\partial M}=\frac{N}{2M} , 
\end{equation}
 Finally after discretization of the derivative we obtain
\begin{equation}
\log{\left(\frac{\Gamma(M+1,N)}{\Gamma(M,N)}\right)}=\frac{N}{2M+1}. \label{dsde}
\end{equation}
This equation is tested in Fig.2, where a half of the inverse of the left hand side,
as obtained from simulations,
is ploted as a function of $\epsilon$. Solid line represents the right hand side 
\footnote{Of course $\epsilon=(M+1/2)/N$ in this case.}.
Similarly to the previous case agreement is very good for $N\geq 12$. It was neseccary
to reduce MC errors to the level of 1\%-3\% to achieve this agreement. Of couse
this test is much more sensitive than the previous one since it requires precise
 measurement of the derivatives.

To conclude, the coincidence method is satisfactory in practice for the number of degrees
of freedom below $\sim 25$. This turns out to be sufficient to see the signatures
of the thermal equilibrium. For more than 12 degrees of fredom the scaling of the
 entropy density is confirmed with the
 accuracy better than 20\% . The saddle point relation $\partial S/\partial E=1/T$
 is also very well reproduced. 

\vspace*{.5cm}
This work is supported in part by the Polish Committee for Scientific
Research under the grants no. 2P03B 08614 and 2P03B 04412.

  \begin{table}    
  \begin{center}
   \begin{tabular}{||c|c||r|r|r|l|l||} \hline\hline
 N & M   &$\nc$ & $\hat{N}_c$ & $\hat{\Gamma}$ & $\hat{\sigma} /\hat{\Gamma} $ &
 $\delta /\Gamma $ \\
\hline\hline
    &    &    4 000  &   218  & 73 376.     & 0.096        & 0.140     \\ 
    &    &    8 000  &  1000  & 63 299.     & 0.045        & 0.007    \\ 
    & 6  &    16 000 &  3866  & 66 214.     & 0.023        & 0.028    \\ 
    &    &    32 000 & 15884  & 64 465.     & 0.011        & 0.001    \\ 
   \cline{2-7}  
 & $ \Gamma $ &\multicolumn{2}{c}{}&\multicolumn{1}{r}{64 416 } & \multicolumn{2}{c||}{}           \\
   \cline{2-7}
    &    & 8 000     & 30    & 2 133 067.  & 0.260        & 0.100        \\ 
    &    & 16 000    & 124   & 2 064 387.  & 0.127        & 0.065       \\ 
 12 & 12 & 32 000    & 516   & 1 984 434.  & 0.062        & 0.024       \\ 
    &    & 64 000    & 2 110 & 1 941 201.  & 0.031        & 0.001       \\
    &    & 128 000   & 8 358 & 1 960 262.  & 0.015        & 0.011       \\ 
   \cline{2-7}  
    & $ \Gamma $ & \multicolumn{2}{c}{}&\multicolumn{1}{r}{1 938 336 }& \multicolumn{2}{c||}{}           \\
   \cline{2-7}
    &    & 20 000   & 4      & 99 995 000. & 0.707        & 0.615           \\ 
    &    & 40 000   & 16     & 99 997 504. & 0.354        & 0.615          \\ 
    & 24 & 80 000   & 106    & 60 376 604. & 0.137        & 0.025          \\ 
    &    & 160 000  & 422    & 60 663 128. & 0.069        & 0.020          \\
    &    & 320 000  & 1596   & 64 160 200. & 0.035        & 0.036          \\ 
   \cline{2-7}  
    & $ \Gamma $ &  \multicolumn{2}{c}{}&\multicolumn{1}{r}{61 903 776 }& \multicolumn{2}{c||}{}           \\
   \hline
    &    & 8 000    & 8   & 7 999 000.  & 0.500        & 0.077             \\ 
    &    & 16 000   & 30  & 8 532 800.  & 0.258        & 0.015          \\ 
    & 6  & 32 000   & 132 & 7 757 334.  & 0.123        & 0.105          \\ 
    &    & 64 000   & 442 & 9 266 824.  & 0.067        & 0.070          \\ 
    &    & 128 000  & 1842& 8 894 610.  & 0.033        & 0.027          \\ 
   \cline{2-7}  
 24 & $ \Gamma $ &  \multicolumn{2}{c}{}&\multicolumn{1}{r}{8 667 720 }& \multicolumn{2}{c||}{}           \\
   \cline{2-7}
    &    & 40 000    & 78    &  99 997 504.  & 0.354    & 0.487           \\ 
    &    & 80 000    & 2100  & 199 997 504.  & 0.250    & 0.025          \\ 
    & 8  & 160 000   & 8334  & 209 834 752.  & 0.128    & 0.076         \\ 
    &    & 320 000   & 33658 & 200 783 680.  & 0.063    & 0.029           \\
    \cline{2-7}  
    & $\Gamma $ & \multicolumn{2}{c}{}&\multicolumn{1}{r}{195 082 320 } & \multicolumn{2}{c||}{}           \\
\hline\hline
   \end{tabular}
  \end{center}
\caption{Monte Carlo results for $\Gamma(M,N)$ (col.5) for different $N$ and $M$.
 The third and fourth column
 give the number of generated configurations $\nc$, and the number of observed
coincidences $\hat{N}_c$. In the last two columns we quote
the Monte Carlo estimate of the relative error ,c.f. Eq.(\ref{err}), and the 
actual relative deviation $\delta/\Gamma=|\hat{\Gamma}-\Gamma|/\Gamma$
 from the exact value $\Gamma$ also quoted in the Table.}
   \end{table}
\newpage

 \begin{figure}[htb]
\epsfig{width=12cm,file=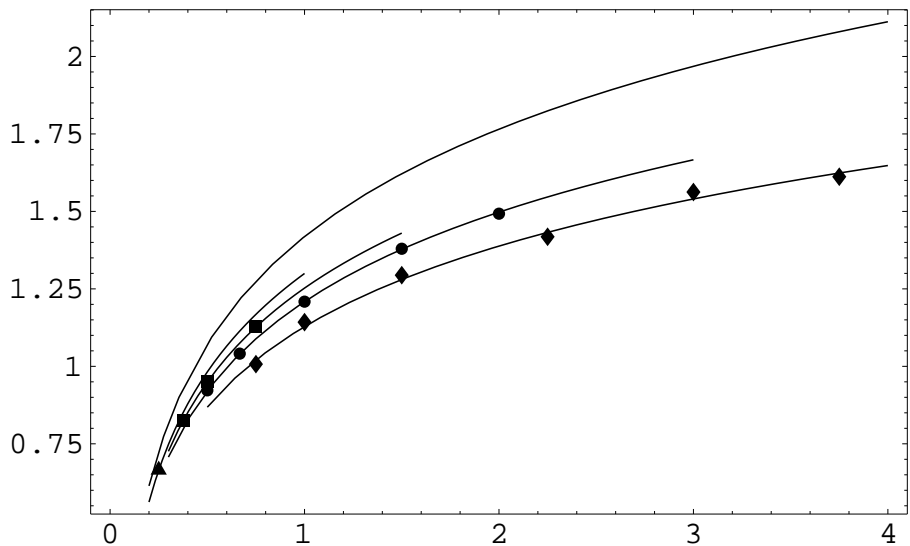}
\caption{Entropy density $s=\frac{1}{N}\log{W(M,N)}$ vs. the energy density 
 $\epsilon=M/N$. Black symbols represent our Monte Carlo results for N=8 (diamonds),
12 (circles), 16 (boxes) and 24 (a triangle). Lower solid lines correspond to the continuum 
approximation, Eq.(\ref{cont}), for each $N$.
 The uppermost solid line represents the scaling,
 thermodynamical limit, Eq.(\ref{thermo}). 
}
\label{fig:f1}
\end{figure}   

 \begin{figure}[htb]
\epsfig{width=12cm,file=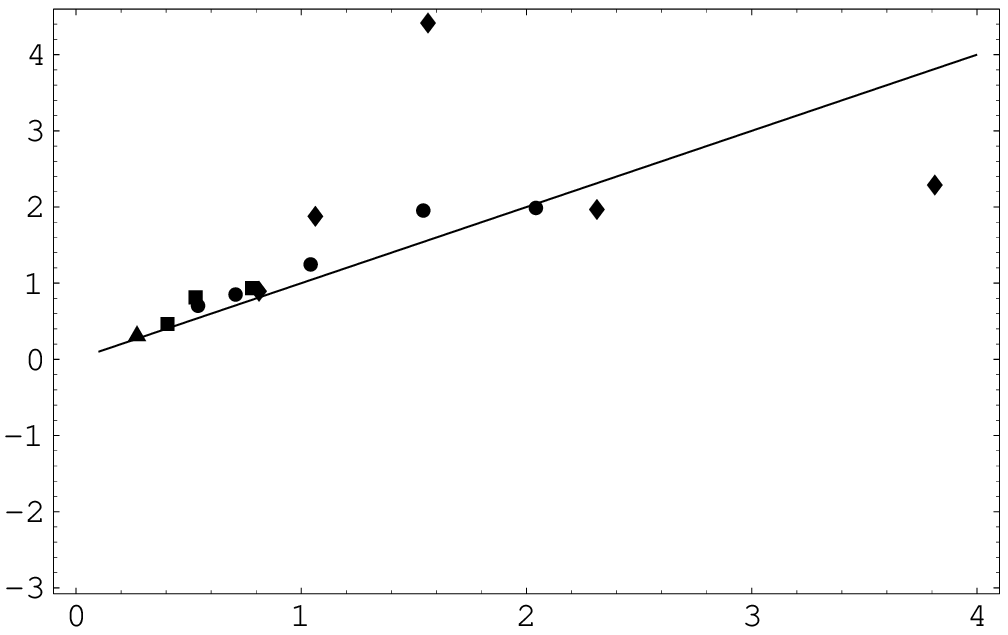}
\caption{Testing the relation \protect{(\ref{dsde})}. Half of the inverse of the finite
 difference (with respect to $M$) of the entropy ${\log \Gamma(M,N)}$, as a function of
 $\protect{\epsilon=M/N}$ for $N=8$ 
(diamonds), 12 (circles), 16(boxes) and 24 (a triangle). Solid line corresponds to the
thermodynamical lmit.}
\label{fig:f2}
\end{figure}


\begin{thebibliography}{99}
\bibitem{la1}
L.Landau and E.Lifszic, Statistical Physics, Chapter XII.
\bibitem{ma1}
S.K. Ma, Statistical Mechanics, World Scientific (1985), p.425.

S.K.Ma, J.Stat.Phys. 26 (1981) 221.

S.K.Ma and M.Payne, Phys. Rev B24 (1981) 3984.
\bibitem{bg} A. Bialas, M. Gazdzicki, Phys. Lett. {\bf B252}(1990)483;
K. Fialkowski, B. Wosiek, and J. Wosiek, Acta Phys. Polon. {\bf
B20} (1989) 639.
\end{thebibliography}
\end{document}